\documentclass[a4paper,11pt]{article}

\usepackage{graphicx}
\usepackage{tikz}
\usepackage{booktabs}
\usepackage{hyperref}
\usepackage{geometry}
\usepackage{xcolor}
\usepackage{array}

\hypersetup{
  colorlinks=true,
  linkcolor=blue,
  urlcolor=blue,
  citecolor=blue
}
\usetikzlibrary{arrows.meta,positioning,fit}

\newcommand{\lblServiceHead}{Voice dialogue service}
\newcommand{\lblServiceBody}{official web client\\(e.g.\ ChatGPT, Grok)}
\newcommand{\lblUserHead}{User}
\newcommand{\lblUserBody}{grants tab-audio sharing\\microphone not captured}
\newcommand{\lblAppNote}{runs locally, no API}
\newcommand{\lblCaptureHead}{Tab audio capture}
\newcommand{\lblCaptureBody}{getDisplayMedia}
\newcommand{\lblAnalysisHead}{Feature extraction}
\newcommand{\lblAnalysisBody}{RMS / voice band / peak}
\newcommand{\lblStateHead}{State generation}
\newcommand{\lblStateBody}{speech, mouth, expression}
\newcommand{\lblAvatarHead}{Rendering layer}
\newcommand{\lblAvatarBody}{mouth and expression,\\gaze and body}
\newcommand{\lblAvatarBackends}{VRM / Live2D / CSS}
\newcommand{\lblAvatarSwap}{interchangeable}
\newcommand{\lblProgressHead}{Continuity support}
\newcommand{\lblProgressBody}{XP / level\\streak / login record}
\newcommand{\lblLogHead}{Practice Log}
\newcommand{\lblLogBody}{phrases and procedures\\saved by date}
\newcommand{\lblOptional}{dashed: optional}
\newcommand{\lblEdgeTabAudio}{tab audio}
\newcommand{\lblEdgeConsent}{consent}

\newcommand{\lblPipeNote}{No speech recognition and no transcription. Only per-frame low-level features are turned into mouth and body motion}
\newcommand{\lblAudioHead}{Tab audio}
\newcommand{\lblAudioBody}{service audio output\\sampled every frame}
\newcommand{\lblFeatHead}{Features}
\newcommand{\lblFeatBody}{RMS / peak / transient\\voice, high band, centroid}
\newcommand{\lblGateHead}{Speech detection}
\newcommand{\lblGateBody}{compared to\\adaptive threshold}
\newcommand{\lblAvStateHead}{Avatar state}
\newcommand{\lblFloorHead}{Noise floor}
\newcommand{\lblFloorBody}{estimated during silence}
\newcommand{\lblSupportHead}{Continuity support}
\newcommand{\lblSupportBody}{XP / level\\streak}
\newcommand{\lblMetricsHead}{Learning metrics}
\newcommand{\lblMetricsBody}{session and speech time\\segment count}
\newcommand{\lblBackendHead}{Rendering backend}
\newcommand{\lblBackendBody}{maps the state to\\3D / 2D / CSS}
\newcommand{\lblEdgeSmooth}{smoothing}
\newcommand{\lblEdgeMap}{state mapping}
\newcommand{\lblEdgeSilence}{silence}
\newcommand{\lblEdgeThreshold}{threshold}
\newcommand{\lblEdgeRecord}{record}
\newcommand{\lblEdgeReward}{reward}

\newcommand{\lblExpServerHead}{Voice AI\\[-1pt]{\scriptsize\mdseries official service or API}}
\newcommand{\lblExpServerBody}{the same AI\\for both conditions}
\newcommand{\lblExpTreatHead}{Treatment}
\newcommand{\lblExpTreatBody}{avatar moves with the AI voice}
\newcommand{\lblExpCtrlHead}{Control}
\newcommand{\lblExpCtrlBody}{no avatar}
\newcommand{\lblExpSubjHead}{Participants}
\newcommand{\lblExpSubjBody}{randomly assigned\\to one screen}
\newcommand{\lblExpTalk}{voice conversation}
\newcommand{\lblExpNote}{The same AI answers behind both screens. Only the screen in front differs; the conversation is not altered.}

\title{An Experimental Apparatus for Embodied Conversational AI:\\[0.6ex]
  {\large A Simple Implementation Coupling a Commercial Voice AI with an Avatar}}
\author{Hiroshi Hamada \and Lin Gu}
\date{November 2026}

\begin{document}
\maketitle

\begin{abstract}
As generative AI becomes integrated into everyday life, understanding how avatars shape human--AI interaction and the perception of AI as a social partner is becoming increasingly important.
  This paper builds a research instrument on audio-output coupling, which connects the audio output of a generative AI to an external avatar. There are two implementations: the browser implementation captures the audio of a tab selected through screen sharing without installing additional software, and the desktop implementation captures the audio of the service embedded in the application. Both work with the voice AI as people ordinarily use it, and as a method neither depends on a particular service.

The implementation reported here, Deskmate, generates speech state, mouth shape, and expression from low-level features such as voice-band mean amplitude, RMS, peak, and transients. Because it uses no microphone, speech recognition, or transcription, Deskmate acquires neither the user's voice nor the content of the conversation. Records of the conversation remain with the voice AI provider, but none are kept automatically on the side of Deskmate. 
This privacy-conscious design minimises additional exposure of conversational data by processing audio locally without automatically storing conversation records. 
Three interchangeable backends are provided for rendering the avatar: 3D (VRM), 2D (Live2D), and CSS.

Deskmate allows researchers to manipulate avatar presentation as an experimental condition, without changing the commercial service's interface, voice, or response generation. Because conversation content is not held fixed, the estimand is the total effect, including the changes in the participant's speech and the AI's responses that the presentation brings about. A server edition, served from the researcher's server for online experiments, is also provided. The effectiveness of avatar presentation has not yet been evaluated; this paper reports the design and implementation.
\end{abstract}

\bigskip


\section{Introduction}

The boundary between people and machines is already thinning in everyday life. Since January 2015, memorial services have been held at Buddhist temples for AIBO robot dogs that can no longer be repaired.\footnote{\raggedright The first through thirteenth services (January 2015 to November 2024) were held at Kofukuji in Isumi, Chiba Prefecture, whose head priest memorialised nearly 900 units over more than ten years. From the fourteenth service (December 11, 2025), the organizer, Aibo Dock, moved the venue to Unmonji in Matsuyama, Ehime Prefecture. nippon.com, Human-robot partnership: Funerals for AIBO (in Japanese), January 9, 2017, \url{https://www.nippon.com/ja/views/b00909/} (accessed September 24, 2026). Aibo Dock, After the 13th Aibo funeral (in Japanese), December 16, 2024, \url{https://www.a-fun.jp/2024/12/16/\%E7\%AC\%AC13\%E5\%9B\%9E\%E3\%82\%A2\%E3\%82\%A4\%E3\%83\%9C\%E8\%91\%AC\%E3\%82\%92\%E7\%B5\%82\%E3\%81\%88\%E3\%81\%A6/} (accessed September 26, 2026). Aibo Dock, The 14th Aibo funeral (in Japanese), February 5, 2026, \url{https://www.a-fun.jp/2026/02/05/\%E7\%AC\%AC14\%E5\%9B\%9E\%E3\%82\%A2\%E3\%82\%A4\%E3\%83\%9C\%E8\%91\%AC/} (accessed September 26, 2026).} In 2025, a woman was reported to have held a wedding ceremony at a venue in Okayama with an AI persona she had created on ChatGPT.\footnote{\raggedright RSK Sanyo Broadcasting, ``I married ChatGPT'' (in Japanese), TBS NEWS DIG, November 8, 2025, \url{https://newsdig.tbs.co.jp/articles/-/2271771} (accessed September 24, 2026).} Embodied AI is also appearing as a product. In July 2025, xAI added a companion feature to the app of its dialogue AI Grok, letting paying subscribers converse with 3D animated characters.\footnote{\raggedright A. Silberling, Elon Musk's Grok is making AI companions, including a goth anime girl, TechCrunch, July 14, 2025, \url{https://techcrunch.com/2025/07/14/elon-musks-grok-is-making-ai-companions-including-a-goth-anime-girl/} (accessed September 24, 2026).} In January 2026, Razer announced an AI companion presented as a 5.5-inch 3D hologram that sits on the desk.\footnote{\raggedright Razer, Razer AVA: 3D Hologram AI Companion, \url{https://www.razer.com/concepts/project-ava} (accessed September 24, 2026).}

Researchers have also examined how people come to treat commercial AI companions as relational partners. Interviews with 18 people who had formed friendships with Replika described a progression from curiosity towards relationships with affective and social value as trust and self-disclosure deepened \cite{skjuve2021}. A mixed-method study of Replika users supported a model in which perceived anthropomorphism and AI authenticity were linked to attachment to the AI through interaction intensity \cite{pentina2023}. Two natural experiments exploiting the removal of Replika's erotic role-play feature and the rollout of GPT-5 further analysed 54,861 posts and 1,452 participants across seven surveys; negative reactions, loss framing, and demands to restore the previous version increased after the updates \cite{defreitas2026}. These studies show that relationships with AI can acquire psychological significance, but they mainly examine accounts from users who formed such relationships or reactions to product changes and do not isolate the effect of presenting a body itself.

One of the questions we need to ask is how an AI with a body, also known as an embodied conversational agent (ECA) \cite{kiuchi2024}, changes people's perception, judgement, and behaviour. Do people attribute a mind to an embodied AI? Does continued interaction give rise to closeness? Do people come to treat AI as a social other rather than a tool? These are empirical questions that experiments can test, and they need to be addressed from the standpoint of the humanities and social sciences. Building a dialogue agent for research, however, means studying something that differs in performance and experience from the commercial services people use every day. Using a commercial service as it is leaves no means of manipulating only the presented body. What is needed is a simple experimental tool that changes only the presented body while leaving the AI people actually use untouched.

Embodied conversational agents have also been investigated in language-learning contexts. Speaking practice in English with conversational agents is associated not only with learning outcomes but with confidence, engagement, motivation, and speaking anxiety \cite{du2024}, and a meta-analysis of comparative studies reports effects on second language learning \cite{lyu2025}. These reviews name embodiment and social presence as design dimensions and argue that conversational agents can function as conversation partners rather than teaching materials \cite{jeon2023,huang2022}. Work treating embodied conversational agents as interactional participants \cite{han2026}, and work manipulating avatar representation to examine its effects on learning experience and speaking anxiety \cite{tan2025,pan2026}, have also appeared. Meng compares combinations of voice modes and instruction conditions within the same voice-dialogue service \cite{meng2026}.

Voice dialogue itself is changing quickly. GPT-Live, released in July 2026, adopts a full-duplex design that handles audio input and output at once and copes with backchannels and interruptions.\footnote{\raggedright OpenAI, Introducing GPT-Live, published July 8, 2026, \url{https://openai.com/index/introducing-gpt-live/} (accessed September 14, 2026).} A voice is a cue for judging that a partner has a mind \cite{schroeder2016}, and the conditions under which people attribute mental states to nonhuman agents have been organised as a theory of anthropomorphism \cite{epley2007}. Yet a four-week randomised controlled trial comparing text with two voice modes found no significant effect of the mode on loneliness or on socialisation with real people \cite{fang2025}. A voice becoming more human-like and a change in people's perception, behaviour, or psychosocial outcomes need to be examined separately. Similarly, the effects of adding visual embodiment to voice interaction need to be examined empirically.

Among platforms for behavioural experiments with large language models, LEXI handles agent configuration, participant registration, questionnaires, and log collection in one place \cite{laban2024}, and Mouakher et al.\ present an environment in which task framing, stakes, and time pressure can be manipulated during a dialogue \cite{mouakher2026}. 
Both focus on text-based interactions with researcher-configured agents, rather than on manipulating the visual presentation of an existing commercial voice AI service. 
LEXI states that its disembodied agent can serve as the control condition in experiments measuring the effects of embodiment.
Open-source tools such as Persona\footnote{\raggedright xikhar, Persona, \url{https://github.com/xikhar/persona} (accessed November 19, 2026).} and AVATAR\footnote{\raggedright arpahls, AVATAR, \url{https://github.com/arpahls/avatar} (accessed November 19, 2026).} already drive a VRM avatar from the audio that a voice AI plays.

The contributions of this paper are threefold. First, we build a research instrument on audio-output coupling, which uses the audio output played by a commercial voice AI as a coupling point to an external system, and realise it in two ways: a browser implementation that captures tab audio and a desktop implementation that captures audio from an embedded frame. Neither modifies the existing service or requires a developer API. The browser implementation worked with two different services without any change to its code. Second, we convert low-level features of the audio into an avatar state independent of rendering technology, so that the same state can be rendered by three backends using 3D, 2D, and CSS. Third, we position as an experimental instrument the ability to manipulate only the avatar presentation while leaving the commercial service's interface, audio, and response generation unchanged, and show how switching presentation conditions lets it be used in randomised experiments. We call the software that implements these contributions Deskmate. What this paper describes is the version with all features as of November 2026 (hereafter the full version); the published demo contains part of it.

\section{Audio-Output Coupling}
\label{sec:coupling}

In this paper, the method of using the audio a service plays as the coupling point to an external system is called audio-output coupling. The browser implementation captures the audio of a tab selected through screen sharing; we call this approach tab-audio coupling. The desktop implementation captures the audio of the service embedded in the application. Neither requires access to the provider's developer API. Audio delivered through such an API can be coupled in the same way, as long as it is played.

The browser implementation uses the audio of a tab the user selects, such as one running ChatGPT's or Grok's voice mode. The avatar responds not to the user's speech but to the audio that the shared tab plays.

The desktop implementation uses Electron, an application platform combining Chromium with Node.js, in which the screen is built with HTML, CSS, and JavaScript but runs as an independent window rather than a browser tab. The desktop implementation opens the dialogue service's page as a frame embedded in that window and captures the audio that frame plays. Because the source is designated by the application, no dialog asking which tab to share ever appears.

In both implementations the video track is stopped immediately on capture and only the audio track is analysed. Processing after capture is shared and is described in Section \ref{sec:impl}. Table \ref{tab:impl} compares the two implementations.

\begin{table}[htp]
  \centering
  \caption{The two implementations compared}
  \label{tab:impl}
  \small
  \begin{tabular}{lll}
    \toprule
    & Browser implementation & Desktop implementation \\
    \midrule
    Audio source & tab chosen by the user & embedded frame \\
    Starting the session & choose a tab each time & one button \\
    Scope of observation & the chosen tab\textsuperscript{*} & the embedded frame \\
    Where the service runs & the user's browser & inside the application \\
    Where credentials live & the user's browser & the application's storage \\
    Microphone captured & no & no \\
    Conversation stored automatically & no (manual saving available) & no (manual saving available) \\
    \bottomrule
  \end{tabular}

  \smallskip
  {\footnotesize \textsuperscript{*}If the whole screen is chosen for sharing and system audio is included, every sound the PC plays.}
\end{table}

In the desktop implementation, the user must log in to the service inside the application, and the login state of the existing browser is not carried over. The desktop implementation states the URL of the page it embeds and the sites the embedded frame is allowed to display. The current desktop implementation is configured for a single service. Because these settings are made per service, several services could be registered in advance and one chosen at launch, but the choice would still be limited to registered services: supporting another service requires writing down, besides its URL, the authentication sites passed through at sign-in and the sites permitted to use the microphone and similar features.

The browser implementation's code contains no processing that depends on a particular service's name, API, DOM structure, or internal protocol: it is enough that the user selects a compatible browser tab that plays audio. The same code can therefore be used with other dialogue services that play audio in a browser. Generality as a method, and the ability to support more services, are common to both; independence, in the sense of connecting the avatar to any service without service-specific code, belongs to the browser implementation alone.

This was confirmed in operation. With the same code, the browser implementation drove the avatar from the voice dialogue of both ChatGPT and Grok (October 2026). Only operation was confirmed; the quality of the mouth movement was not compared across services. Thresholds such as those for detecting speech were tuned on one service's voice, so a service whose voice differs in character may not behave in the same way.

Some dialogue services play their audio outside any browser tab. The voice of Gemini running in Chrome's side panel does not belong to a tab and cannot be captured as tab audio. We therefore added to the browser implementation an option that also captures system audio when the whole screen is chosen for sharing. The option is a standard browser feature, not processing written for a particular service. With it, the avatar could also be driven by Gemini's voice dialogue. What is captured in this case, however, is every sound the PC plays. Notification sounds and the audio of other applications enter the avatar's responses as well, and the property of limiting observation to the chosen tab is lost. The microphone is still not opened, analysis still completes within the browser, and audio is still not sent outside. For services whose tab audio suffices, the tab should be chosen.

Deskmate does not directly access the user's microphone. The captured AI audio is only converted into features within the page that analyses it, and is not sent outside that page, not even to the local server. Driving the avatar needs only the AI audio the service plays, not the user's speech, so no call that acquires the microphone is implemented. Audio analysis computes low-level features including RMS, voice-band mean amplitude, amplitude envelope, and transients within the browser, and does not automatically store the waveform or a transcript. A purpose that evaluates the user's own speech, such as gains in pronunciation, would however require a separate input path.

\section{Properties as an Experimental Instrument}
\label{sec:instrument}

Building a dialogue system for research allows content and presentation to be controlled, but the result may differ in performance and feel from the commercial services people ordinarily use. Using a commercial service is closer to actual use, but its internals cannot be changed. Deskmate leaves the commercial service's interface, voice, and response generation unchanged and allows avatar presentation to be manipulated independently. 
Table \ref{tab:tools} shows how it differs from existing experimental platforms.

\begin{table}[tbp]
  \centering
  \caption{An existing experimental platform, a system built on a developer API, and Deskmate}
  \label{tab:tools}
  \small
  \begin{tabular}{l*{3}{>{\raggedright\arraybackslash}p{3.2cm}}}
    \toprule
    & LEXI \cite{laban2024} & Built on a developer API & Deskmate \\
    \midrule
    Modality & text & voice possible & voice \\
    Partner & configured by the researcher & configured by the researcher & the official service itself \\
    Content & set by prompts & set by prompts & not intervened in \\
    Presented body & none & must be built separately & main object of manipulation \\
    Extra usage fees & depend on the model used & usage-based & no fee added by Deskmate \\
    Handling of conversation logs & collected by the platform & collected by the builder & not stored automatically (manual saving available) \\
    \bottomrule
  \end{tabular}
\end{table}

A developer voice API allows voice and content to be controlled. The interface, the segmentation of turns, the handling of interruptions, and the voice settings, however, must then be implemented by the researcher, and commercial usage-based APIs incur fees. With Deskmate, participants use the official service as usual, and the experimenter changes only the avatar presentation.

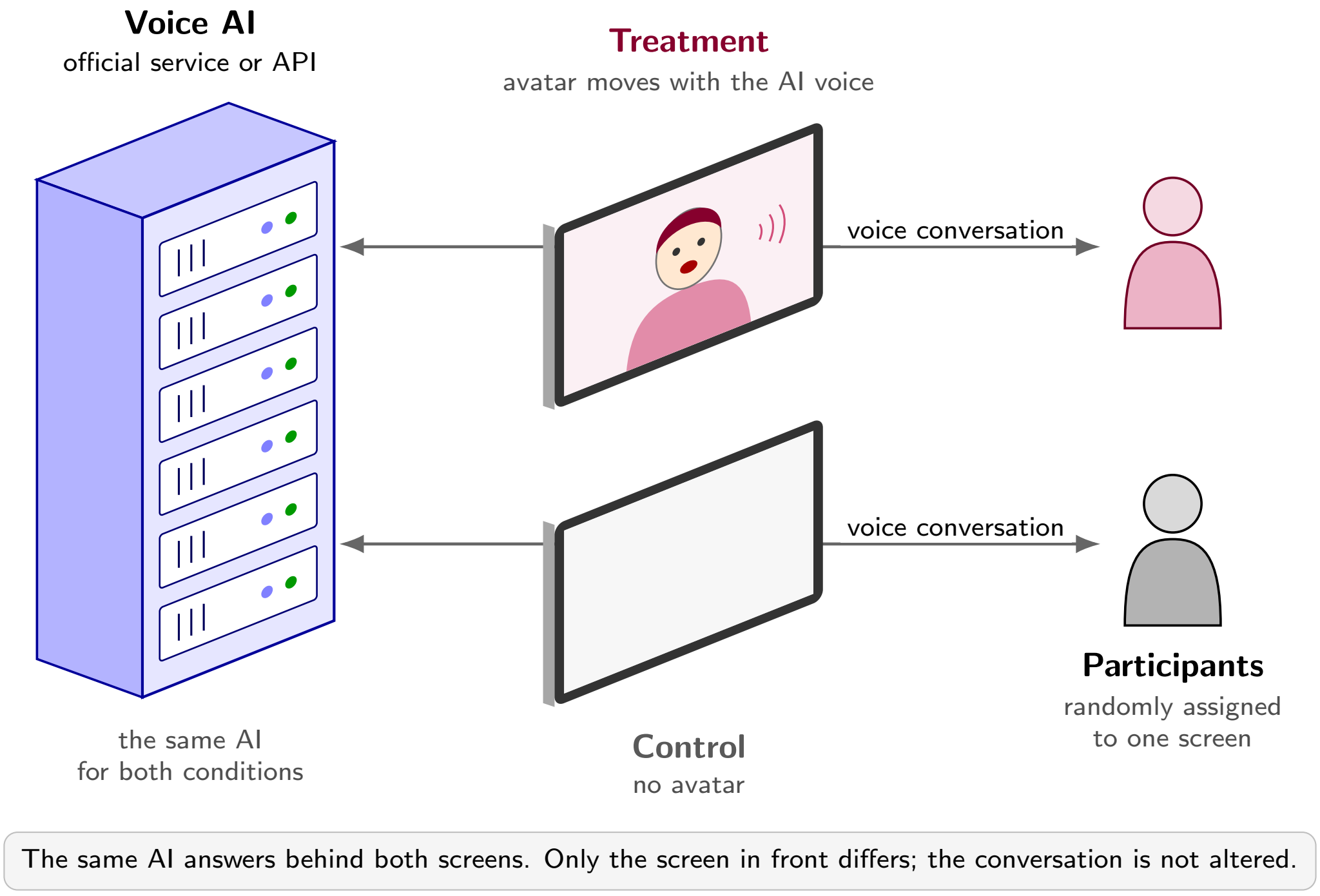
\begin{figure}[tbp]
  \centering
  \resizebox{0.85\linewidth}{!}{\begin{tikzpicture}[
  x=1cm,
  y=1cm,
  font=\sffamily\small,
  face/.style={cm={1,0.4,0,1,(0,0)}},
  talk/.style={{Latex[length=2.6mm]}-{Latex[length=2.6mm]}, line width=0.9pt, draw=black!60},
  tag/.style={font=\sffamily\scriptsize, fill=white, inner sep=2pt},
  head/.style={font=\sffamily\small\bfseries, align=center},
  sub/.style={font=\sffamily\scriptsize, align=center, text=black!70}
]
  \draw[talk] (2.05,3.85) -- node[tag, above, pos=0.81] {\lblExpTalk} (10.0,3.85);
  \draw[talk] (2.05,0.75) -- node[tag, above, pos=0.81] {\lblExpTalk} (10.0,0.75);

  \def\sw{2.0}\def\sh{5.0}
  \begin{scope}[shift={(0,-0.85)}]
    \fill[blue!22, draw=blue!60!black, line width=0.7pt]
      (0,\sh) -- (\sw,\sh+0.4*\sw) -- ++(-1.1,0.4) -- (-1.1,\sh+0.4) -- cycle;
    \fill[blue!30, draw=blue!60!black, line width=0.7pt]
      (0,0) -- (0,\sh) -- (-1.1,\sh+0.4) -- (-1.1,0.4) -- cycle;
    \begin{scope}[face]
      \fill[blue!10, draw=blue!60!black, line width=0.8pt] (0,0) rectangle (\sw,\sh);
      \foreach \k in {0,...,5} {
        \pgfmathsetmacro{\yb}{0.3+\k*0.76}
        \fill[white, draw=blue!45!black, line width=0.5pt, rounded corners=1pt]
          (0.18,\yb) rectangle (\sw-0.18,\yb+0.56);
        \foreach \j in {0,1,2} {
          \draw[blue!35!black, line width=0.6pt] (0.38+\j*0.13,\yb+0.14) -- ++(0,0.28);
        }
        \fill[green!60!black] (\sw-0.45,\yb+0.28) circle (0.06);
        \fill[blue!50] (\sw-0.70,\yb+0.28) circle (0.06);
      }
    \end{scope}
  \end{scope}
  \node[head, anchor=south] at (0.5,5.55) {\lblExpServerHead};
  \node[sub, anchor=north] at (0.5,-1.05) {\lblExpServerBody};

  \begin{scope}[shift={(4.3,2.15)}]
    \fill[black!35] (0,0) -- (0,1.9) -- (-0.12,1.94) -- (-0.12,0.04) -- cycle;
    \begin{scope}[face]
      \fill[black!80, rounded corners=3pt] (0,0) rectangle (2.8,1.9);
      \fill[white, opacity=0.9] (0.1,0.1) rectangle (2.7,1.8);
      \fill[purple!6] (0.1,0.1) rectangle (2.7,1.8);
      \fill[purple!45] (0.75,0.1) .. controls (0.8,0.62) and (1.05,0.72) .. (1.4,0.72)
        .. controls (1.75,0.72) and (2.0,0.62) .. (2.05,0.1) -- cycle;
      \fill[orange!18, draw=black!55, line width=0.5pt] (1.4,1.12) ellipse (0.34 and 0.40);
      \fill[purple!70!black] (1.06,1.2) .. controls (1.05,1.62) and (1.75,1.62) .. (1.74,1.2)
        .. controls (1.6,1.42) and (1.2,1.42) .. cycle;
      \fill[black!80] (1.27,1.14) circle (0.04) (1.53,1.14) circle (0.04);
      \fill[red!65!black] (1.4,0.93) ellipse (0.09 and 0.06);
      \foreach \r in {0.16,0.28,0.40} {
        \draw[purple!70, line width=0.6pt] ([shift={(-30:\r)}]2.0,1.0) arc (-30:30:\r);
      }
    \end{scope}
  \end{scope}
  \node[head, text=purple!70!black, anchor=south] at (5.7,5.75) {\lblExpTreatHead};
  \node[sub, anchor=south] at (5.7,5.35) {\lblExpTreatBody};

  \begin{scope}[shift={(4.3,-0.95)}]
    \fill[black!35] (0,0) -- (0,1.9) -- (-0.12,1.94) -- (-0.12,0.04) -- cycle;
    \begin{scope}[face]
      \fill[black!80, rounded corners=3pt] (0,0) rectangle (2.8,1.9);
      \fill[white, opacity=0.9] (0.1,0.1) rectangle (2.7,1.8);
      \fill[black!3] (0.1,0.1) rectangle (2.7,1.8);
    \end{scope}
  \end{scope}
  \node[head, text=black!65, anchor=north] at (5.7,-1.1) {\lblExpCtrlHead};
  \node[sub, anchor=north] at (5.7,-1.55) {\lblExpCtrlBody};

  \foreach \y/\c in {3.85/purple, 0.75/black} {
    \begin{scope}[shift={(10.75,\y)}]
      \fill[\c!30, draw=\c!60!black, line width=0.7pt]
        (-0.5,-0.85) .. controls (-0.5,-0.1) and (-0.3,0.05) .. (0,0.05)
        .. controls (0.3,0.05) and (0.5,-0.1) .. (0.5,-0.85) -- cycle;
      \fill[\c!15, draw=\c!60!black, line width=0.7pt] (0,0.42) circle (0.3);
    \end{scope}
  }
  \node[head, anchor=north] at (10.75,-0.25) {\lblExpSubjHead};
  \node[sub, anchor=north] at (10.75,-0.7) {\lblExpSubjBody};

  \node[draw=black!25, fill=black!4, rounded corners=4pt, font=\sffamily\scriptsize,
        align=center, inner sep=5pt, anchor=north] at (5.4,-2.25) {\lblExpNote};
\end{tikzpicture}}
  \caption{Concept of Deskmate as an experimental instrument. The same voice AI answers behind both screens. It may be the official service used as usual or a model reached through a developer API; in either case, Deskmate uses only the voice that is played. Participants are randomly assigned to one screen: the treatment screen shows an avatar that moves with the AI's voice, and the control screen shows no avatar.}
  \label{fig:apparatus}
\end{figure}

When Deskmate is launched for an experiment, the condition can be specified by a URL parameter or by a start-up option of the desktop implementation: for example, no avatar, the same model shown in a still state, or the mouth synchronised with the audio. In every condition, Deskmate leaves the dialogue service's interface, audio, and responses untouched. Synchronisation here means mouth movement based on low-level audio features, not phoneme recognition. Figure \ref{fig:apparatus} illustrates the idea with two conditions.

Served over HTTPS from the researcher's server, the browser implementation can also be used in online experiments. Participants open a link and share the dialogue service's tab on their own device; nothing needs to be installed. Audio analysis completes within the participant's browser, so even when served from a server, the audio the participant hears is not itself sent to the researcher. In addition, since the panels in the side pane can be replaced according to purpose, a questionnaire can be shown there to obtain participants' responses before and after the dialogue. Records such as usage time and the AI's speaking time, together with questionnaire responses, can thus be collected from many participants without bringing them into a laboratory. For this use we implemented a server edition that adds condition assignment and the receipt of records; its configuration is described in Section \ref{sec:server}.

This design also has practical advantages over an alternative in which the researcher runs speech recognition, a language model, and speech synthesis locally and serves the dialogue from a laboratory machine. First, many participants can take part at the same time. Speech generation runs on the commercial service for each participant, so the number of concurrent sessions is not limited by the researcher's computing resources. Second, the server can be a low-specification machine. It only delivers static files and receives records; audio analysis and rendering run in each participant's browser, and no GPU is required. Third, the avatar is added to the voice AI at its current best performance. Models that can run on a single local machine generally fall short of the latest commercial voice services in response quality, latency, and naturalness of speech, and an effect of the avatar measured with a weaker conversational partner may not carry over to the services people actually use.

Leaving the dialogue to a commercial service has its drawbacks. The experimenter does not control what is said in the conversation. If the presentation changes the participant's stance, the participant's speech changes, and so does the service's reply. The content of the conversation lies downstream of the treatment as a post-treatment variable, and what is estimated is the total effect that includes this process. Adding a post-treatment quantity such as how interesting the conversation was as a covariate in a regression destroys the comparability that randomisation secured \cite{montgomery2018}. Estimating the direct effect of the body requires fixing the content at the design stage, for example by presenting recorded audio to all participants. Deskmate itself records no conversation, so what was said cannot be analysed; where logs are needed, they must be obtained from the commercial service with consent. Finally, results are tied to the service at the time of the study. Because provider updates change the character of the voice and the replies, the time of the study and the version of the service need to be recorded.

\section{System Architecture and Implementation}
\label{sec:impl}

\begin{figure}[tbp]
  \centering
  \resizebox{\linewidth}{!}{\begin{tikzpicture}[
  x=1cm,
  y=1cm,
  font=\sffamily\small,
  box/.style={draw=#1, fill=#1!8, rounded corners=4pt, line width=0.8pt, align=center, minimum width=4.2cm, minimum height=1.2cm, inner sep=5pt},
  widebox/.style={draw=#1, fill=#1!8, rounded corners=4pt, line width=0.8pt, align=center, minimum width=5.2cm, minimum height=1.35cm, inner sep=5pt},
  group/.style={draw=black!25, fill=black!3, rounded corners=6pt, line width=0.7pt},
  arrow/.style={-{Latex[length=3mm]}, line width=0.9pt, draw=black!65},
  softarrow/.style={-{Latex[length=3mm]}, line width=0.75pt, draw=black!45, dashed},
  tag/.style={font=\sffamily\scriptsize, fill=white, inner sep=2pt},
  note/.style={draw=black!25, fill=black!4, rounded corners=4pt, align=center, font=\sffamily\small, inner sep=5pt}
]
  \node[widebox=blue, text width=4.8cm] (chatgpt) at (0,2.8)
    {\textbf{\lblServiceHead}\\\lblServiceBody};

  \node[box=gray, text width=3.8cm] (user) at (0,-1.3)
    {\textbf{\lblUserHead}\\\lblUserBody};

  \node[group, minimum width=5.8cm, minimum height=6.5cm] (localapp) at (6.2,0.75) {};
  \node[font=\sffamily\bfseries\large] at (6.2,3.65) {Deskmate};
  \node[font=\sffamily\scriptsize] at (6.2,3.25) {\lblAppNote};

  \node[box=green, text width=4.0cm] (capture) at (6.2,2.1)
    {\textbf{\lblCaptureHead}\\\lblCaptureBody};
  \node[box=green, text width=4.0cm] (analysis) at (6.2,0.75)
    {\textbf{\lblAnalysisHead}\\\lblAnalysisBody};
  \node[box=green, text width=4.0cm] (state) at (6.2,-0.6)
    {\textbf{\lblStateHead}\\\lblStateBody};

  \node[box=purple, text width=4.0cm, minimum height=1.6cm] (avatar) at (12.7,2.8)
    {\textbf{\lblAvatarHead}\\\lblAvatarBody\\{\scriptsize \lblAvatarBackends}\\[-1pt]{\scriptsize \lblAvatarSwap}};
  \node[box=orange, dashed, text width=4.0cm] (progress) at (12.7,0.75)
    {\textbf{\lblProgressHead}\\\lblProgressBody};
  \node[box=orange, dashed, text width=4.0cm] (log) at (12.7,-1.3)
    {\textbf{\lblLogHead}\\\lblLogBody};
  \node[font=\sffamily\scriptsize, text=black!55, anchor=north east, inner sep=1pt] at ([yshift=-3pt]log.south east) {\lblOptional};

  \draw[arrow] (chatgpt.east) -- node[tag, above] {\lblEdgeTabAudio} (capture.west);
  \draw[softarrow] (user.east) -- node[tag, below] {\lblEdgeConsent} (analysis.west);
  \draw[arrow] (capture.south) -- (analysis.north);
  \draw[arrow] (analysis.south) -- (state.north);

  \coordinate (split) at (9.55,-0.6);
  \draw[arrow] (state.east) -- (split) |- (avatar.west);
  \draw[softarrow] (split) |- (progress.west);
  \draw[softarrow] (split) |- (log.west);
\end{tikzpicture}}
  \caption{Architecture of the browser implementation of Deskmate. The voice-dialogue service runs in its official UI; Deskmate captures only the tab audio in the browser and performs avatar control and general knowledge-acquisition support locally.}
  \label{fig:architecture}
\end{figure}
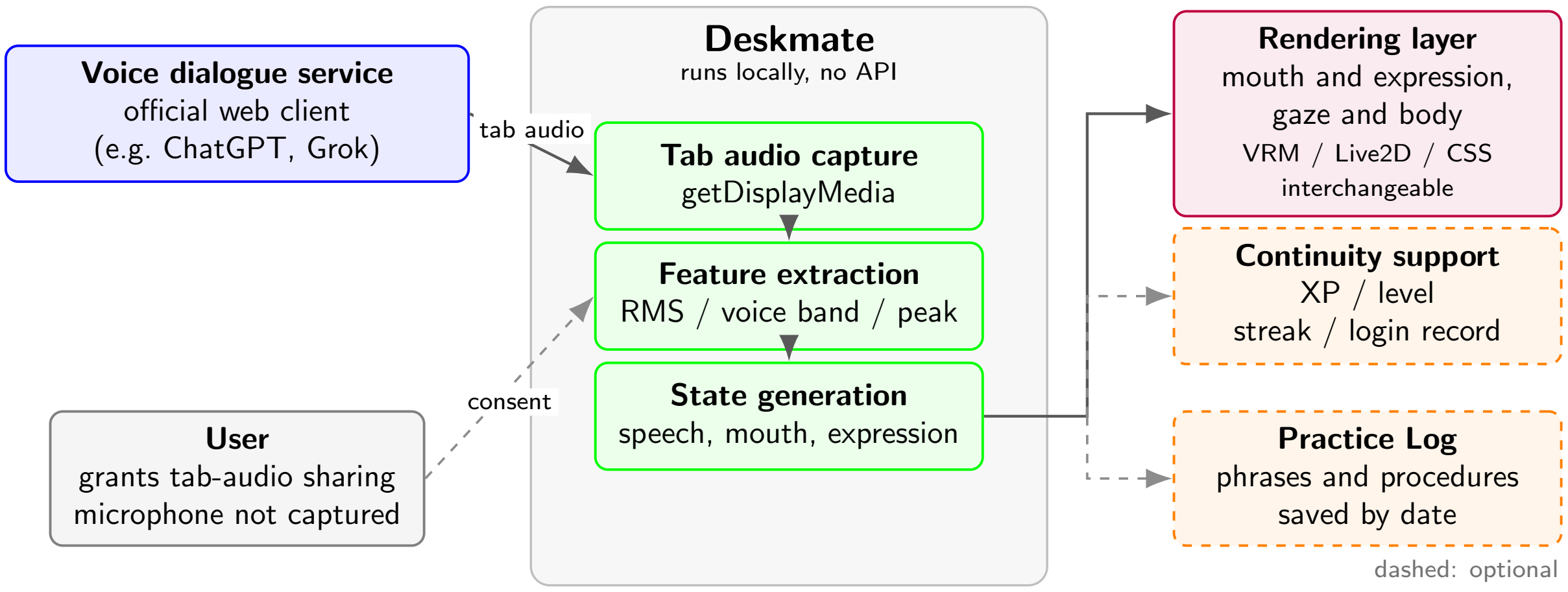

The browser implementation consists of a local Node.js server and a browser client. The server provides static files, the avatar model file, and an API for saving user-selected phrases. The client is implemented in HTML, CSS, and JavaScript, and passes an avatar state computed from local audio features to the rendering layer. Figure \ref{fig:architecture} shows the architecture of the browser implementation.

\texttt{server.mjs} serves the static files and the model, and appends text the user has asked to save to \texttt{useful\_phrases.md}. Audio capture, feature computation, speech detection, and avatar state generation complete within the browser, and audio is not sent to the server. A local server is used so that text can be appended to a file from the browser and the model file can be loaded from the same origin.

\begin{figure}[tbp]
  \centering
  \includegraphics[width=\linewidth]{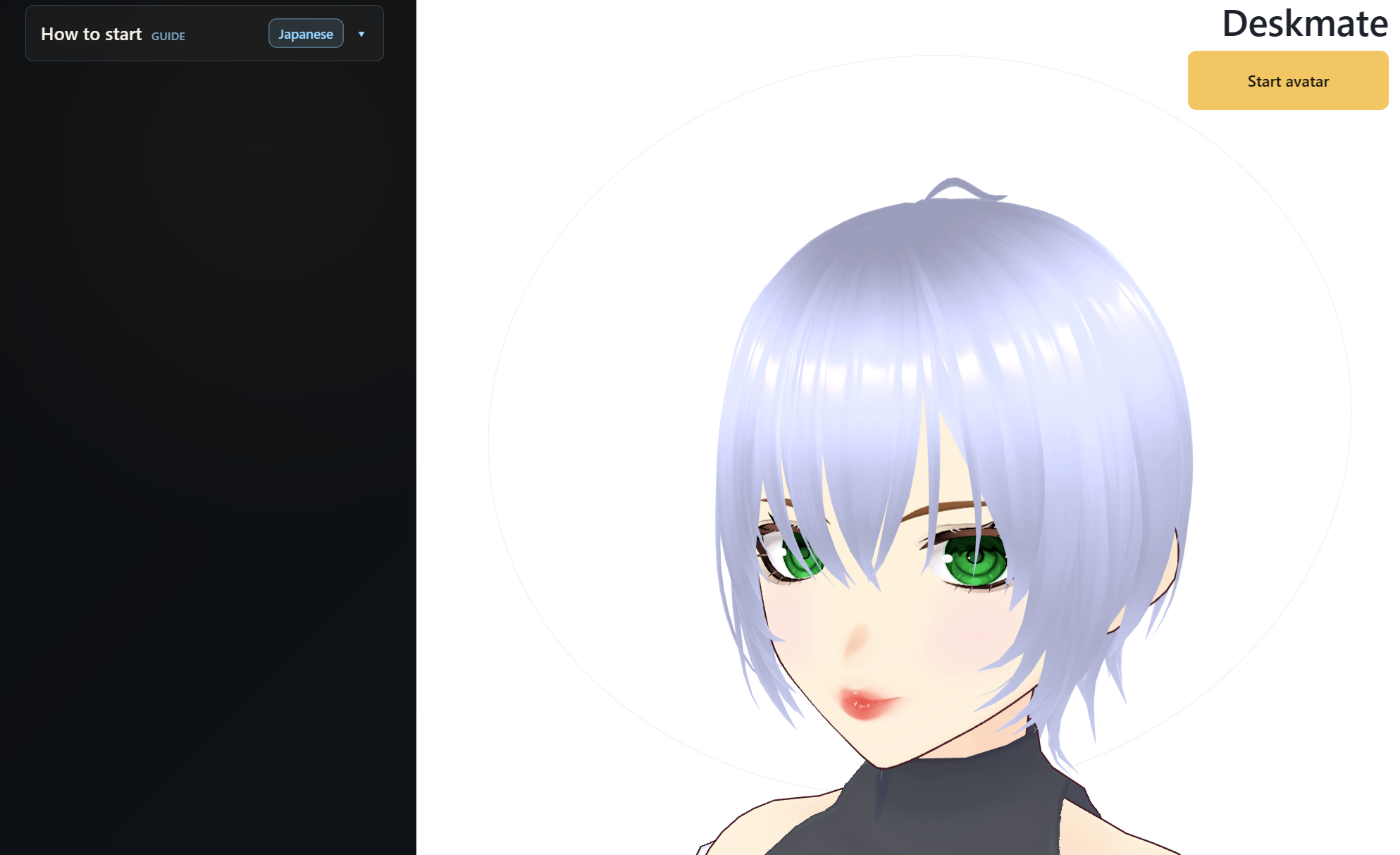}
  \caption{The screen of the published demo, displayed in English. On the right is the stage area drawn by the 3D backend; on the left is the guide to getting started, shown collapsed. The panels can be replaced according to purpose; in ordinary use they hold the learning-support panels. The avatar faces at an angle because of its idle turn.}
  \label{fig:ui}
\end{figure}

The browser client consists of \texttt{index.html}, \texttt{styles.css}, and \texttt{app.js} under \texttt{public/}. The screen has a stage area displaying the avatar and a side pane holding panels for level and consecutive days of use, conversation topics, review of previously saved phrases, and phrase saving. What each panel does is described below. Drawing in the avatar area is the responsibility of a rendering backend that takes the avatar state produced by the audio analysis as input. Backends are interchangeable; their configuration is described in Section \ref{sec:backends}. Figure \ref{fig:ui} shows this screen in the published demo: the learning-support panels are replaced by a guide to getting started, and the stage is given more room.

The browser implementation in the full version runs as a local Node.js application. It assumes Node.js 20 or later and has no npm dependency at run time. All libraries needed for rendering are loaded by the browser from a CDN. A user who wants a different VRM model, or a Live2D sample model for the 2D backend, obtains it from its distributor or makes it themselves. Assets differ in their terms of use, and users are responsible for checking the applicable licensing terms. The published demo differs from the full version. It serves the same client as static pages without a server; it omits the panels for saving and reviewing phrases, which depend on the server, and the 2D backend, and fixes the 3D model to a single one. The full version is available from the author on request for research purposes only.

\subsection{Interchangeable Rendering Backends}
\label{sec:backends}

The avatar state is a tuple of seven values: whether speech is in progress, volume, energy, intensity, mouth opening, mouth shape, and expression. It contains no term from any rendering technology. A rendering backend receives this tuple together with gaze and blinking, which are generated independently of the audio, and nothing else; the audio analysis upstream does not know which backend is running. The current implementation has the three shown in Table \ref{tab:backends}, each belonging to a different technology family.

\begin{table}[tbp]
  \centering
  \caption{The three rendering backends}
  \label{tab:backends}
  \small
  \begin{tabular}{lllp{4.6cm}}
    \toprule
    Setting & Implementation & Asset required & Intended use \\
    \midrule
    \texttt{vrm} & three.js + three-vrm & VRM model & Where a 3D idiom is wanted \\
    \texttt{live2d} & PixiJS + Cubism & Cubism model & Where a 2D idiom is wanted \\
    \texttt{css} & CSS only & none & Where no external resource is available \\
    \bottomrule
  \end{tabular}
\end{table}

Switching is done with a control in the corner of the stage, and the choice is stored in the browser. A setting given in the URL query string takes precedence over the stored value, so a single link can invoke a particular backend without disturbing the stored choice. In the current implementation, switching reloads the page.

\begin{figure}[tbp]
  \centering
  \begin{minipage}[t]{0.48\linewidth}
    \centering
    \includegraphics[width=\linewidth]{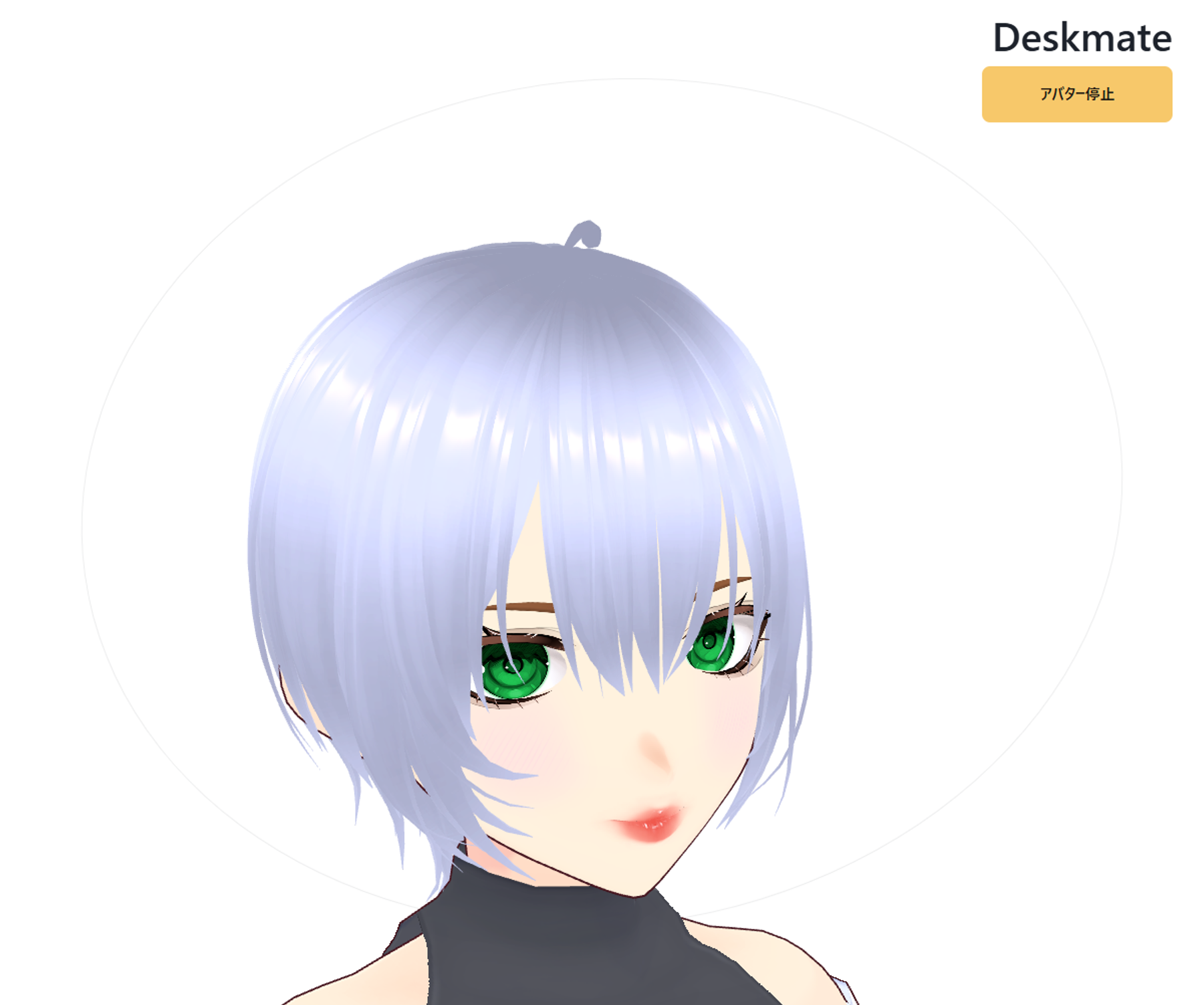}\\
    (a) \texttt{vrm}
  \end{minipage}\hfill
  \begin{minipage}[t]{0.48\linewidth}
    \centering
    \includegraphics[width=\linewidth]{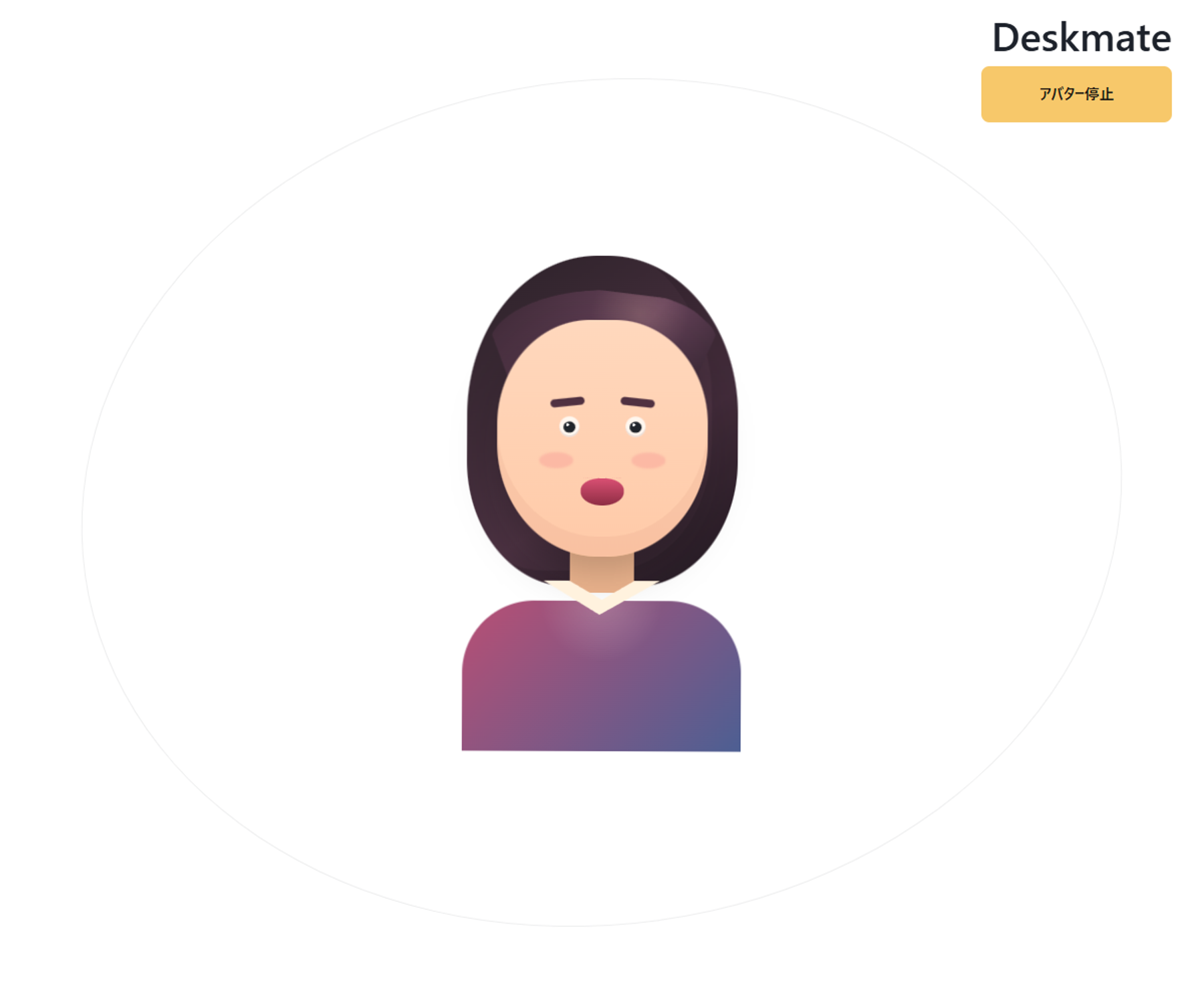}\\
    (b) \texttt{css}
  \end{minipage}
  \caption{Two rendering backends consuming the same avatar state. (a) is a VRM model created by the first author, drawn with three-vrm; (b) is drawn with CSS alone. Both are cropped from the online-experiment screen. The difference lies only on the rendering side; the path from tab-audio capture to state generation is shared.}
  \label{fig:backends}
\end{figure}

The 3D and 2D backends accept user-supplied models. Although the three backends differ in resources, coordinate systems, and rendering parameters, they consume the same avatar state and require no change to audio analysis. The 3D backend represents gaze and body orientation, the 2D backend can use existing Cubism assets, and the CSS backend requires no external model asset. Figure \ref{fig:backends} shows the 3D and CSS backends\footnote{The copyright of VRoid belongs to pixiv.}.

\subsection{Audio Response and Knowledge-Acquisition Support}
\label{sec:audio}

\begin{figure}[tbp]
  \centering
  \resizebox{\linewidth}{!}{\begin{tikzpicture}[
  x=1cm,
  y=1cm,
  font=\sffamily\small,
  box/.style={draw=#1, fill=#1!8, rounded corners=4pt, line width=0.8pt, align=center, minimum width=3.6cm, minimum height=1.5cm, inner sep=5pt},
  smallbox/.style={draw=#1, fill=#1!8, rounded corners=4pt, line width=0.7pt, align=center, minimum width=3.2cm, minimum height=1.0cm, inner sep=4pt},
  arrow/.style={-{Latex[length=3mm]}, line width=0.9pt, draw=black!65},
  loop/.style={-{Latex[length=2.6mm]}, line width=0.75pt, draw=black!45},
  softarrow/.style={-{Latex[length=3mm]}, line width=0.75pt, draw=black!45, dashed},
  tag/.style={font=\sffamily\scriptsize, fill=white, inner sep=2pt, align=center},
  note/.style={draw=black!25, fill=black!4, rounded corners=4pt, align=center, font=\sffamily\small, inner sep=5pt}
]
  \node[note, text width=13.0cm] (note) at (7.4,3.8)
    {\lblPipeNote};

  \node[box=blue, text width=3.6cm] (audio) at (0,1.5)
    {\textbf{\lblAudioHead}\\\lblAudioBody};
  \node[box=green, text width=4.0cm] (features) at (4.8,1.5)
    {\textbf{\lblFeatHead}\\\lblFeatBody};
  \node[box=orange, text width=3.7cm] (gate) at (9.5,1.5)
    {\textbf{\lblGateHead}\\\lblGateBody};
  \node[box=purple, text width=4.0cm] (state) at (16.4,1.5)
    {\textbf{\lblAvStateHead}\\speaking / mouthOpen\\voiceShape / expression};

  \node[smallbox=orange, text width=3.6cm] (floor) at (9.5,-1.1)
    {\textbf{\lblFloorHead}\\\lblFloorBody};

  \node[box=orange, text width=3.4cm] (support) at (0,-3.4)
    {\textbf{\lblSupportHead}\\\lblSupportBody};
  \node[box=green, text width=4.0cm] (metrics) at (4.8,-3.4)
    {\textbf{\lblMetricsHead}\\\lblMetricsBody};
  \node[box=red, text width=4.0cm] (params) at (16.4,-3.4)
    {\textbf{\lblBackendHead}\\\lblBackendBody};

  \draw[arrow] (audio.east) -- (features.west);
  \draw[arrow] (features.east) -- (gate.west);
  \draw[arrow] (gate.east) -- node[tag] {\lblEdgeSmooth} (state.west);
  \draw[arrow] (state.south) -- node[tag, right] {\lblEdgeMap} (params.north);
  \draw[loop] ([xshift=-1.1cm]gate.south) -- node[tag, left] {\lblEdgeSilence} ([xshift=-1.1cm]floor.north);
  \draw[loop] ([xshift=1.1cm]floor.north) -- node[tag, right] {\lblEdgeThreshold} ([xshift=1.1cm]gate.south);
  \draw[arrow] ([xshift=-2.0cm]gate.south) -- ++(0,-0.5) -| node[tag, pos=0.78, right] {\lblEdgeRecord} (metrics.north);
  \draw[arrow] (metrics.west) -- node[tag, above] {\lblEdgeReward} (support.east);
  \draw[softarrow] (note.south) -- (features.north);
\end{tikzpicture}}
  \caption{The pipeline from audio analysis to avatar control and continuity support. Speech activity and visual response are generated from audio features, not from the content of what is said.}
  \label{fig:pipeline}
\end{figure}
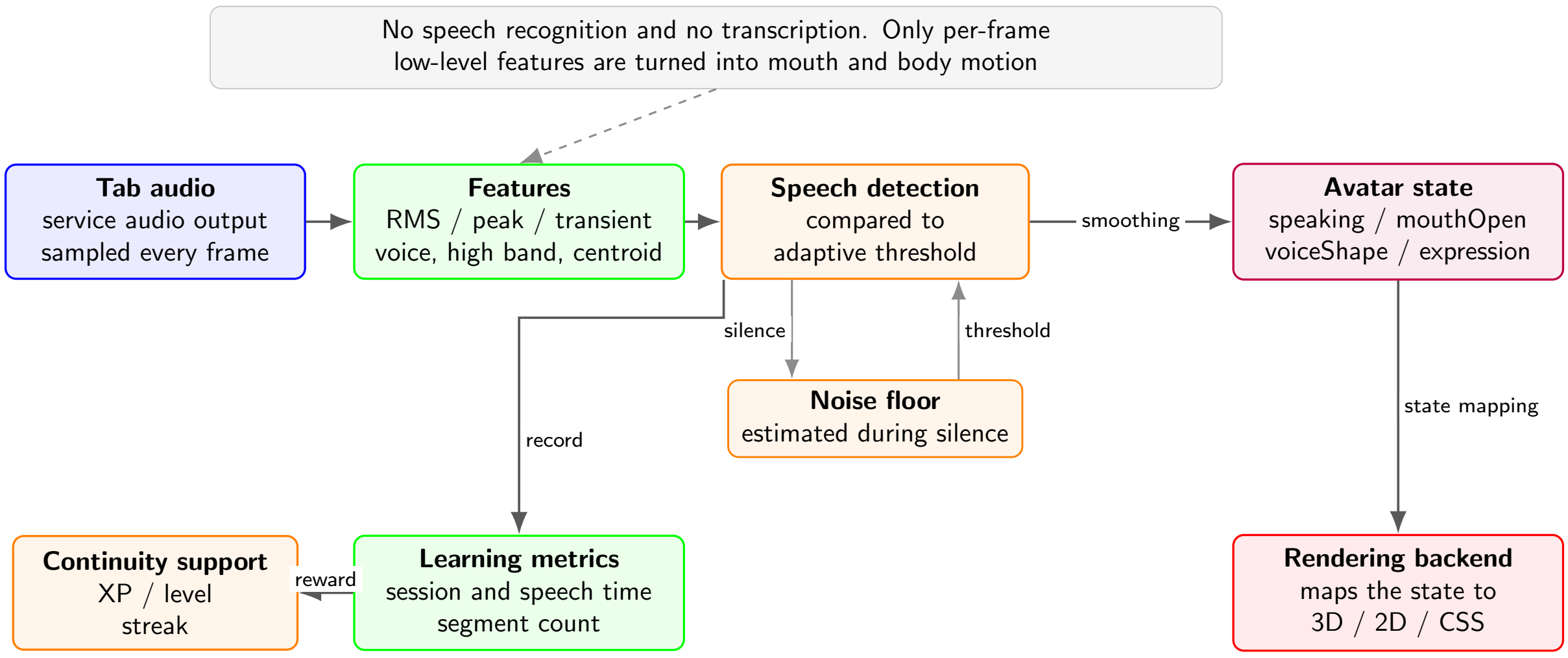

Audio analysis uses \texttt{AudioContext} and \texttt{AnalyserNode} from the Web Audio API. Here, low-level features are short-window numerical descriptions of waveform amplitude and change, rather than the meaning of the words spoken. Voice-band mean amplitude is the mean of the amplitudes of frequency components between 120 and 4200 Hz, the range designated as the voice band in this implementation. RMS is the root mean square of the waveform amplitudes in a short window, whereas peak is the maximum amplitude in that window. For example, two windows can have similar RMS values while the one containing a single large-amplitude sample has a higher peak. A transient is a value representing a sudden rise in amplitude. Each frame also computes high-band mean amplitude and spectral centroid. Figure \ref{fig:pipeline} shows the flow from feature extraction through avatar state and rendering parameters to learning metrics. We call this one-directional flow the pipeline: each stage takes only the output of the stage before it, and whatever is not passed on does not reach any point beyond.

Speech detection compares the audio level with a threshold that follows the noise level, which is estimated continuously during silence.

Mouth opening is computed from audio features, smoothed so that it does not jump, and passed to the rendering layer as part of the avatar state. Each backend maps this value to its own representation of the mouth.

Expression takes the states \texttt{neutral}, \texttt{happy}, \texttt{thinking}, \texttt{surprise}, and \texttt{concerned}. These do not analyse the meaning of what is said; they are a heuristic that visualises the rhythm of the conversation from audio features.

The side pane can display learning-support panels suited to the purpose of use. Possible panels include, for example, a display of experience points and consecutive days of use based on usage time and amount of speech, a panel suggesting conversation topics, and a review panel showing expressions the user has saved manually; the current implementation provides these as examples. The effects of such elements that encourage continued use have been examined in prior work \cite{sailer2020,shortt2023}. The panels are independent of avatar state and can be replaced according to purpose. None of them transcribes or analyses the meaning of speech, and none evaluates learning attainment.

\subsection{Server Edition for Online Experiments}
\label{sec:server}

The server edition is the browser implementation served from the researcher's server; it is not a third coupling method. The path from audio capture to rendering is the same as in the browser implementation and runs in the participant's browser. What the server edition adds is threefold: condition assignment, a study procedure, and the receipt of records.

\texttt{experiment/server.mjs} wraps the handling of the ordinary server and adds assignment and the receipt of submissions. The researcher runs it on their own server and serves it over HTTPS, because tab-audio sharing works only in a secure context. Each participant receives a link containing a participant ID; the ID issued by a recruitment service can be used as it is. The server picks a condition at random from among those with the fewest participants assigned so far, and always returns the same condition for the same ID. The name of the condition appears neither in the URL nor on the screen.

Participants proceed through information and consent, a pre-questionnaire, a connection check, the dialogue, and a post-questionnaire, and finally submit their records and responses and receive a completion code. In the connection check, the participant shares the dialogue service's tab, and cannot proceed to the dialogue until Deskmate has received the AI's voice for a set number of seconds. The avatar is shown only during the dialogue. It is hidden during the pre-questionnaire and the connection check, so that the treatment does not leak into the baseline measurement, and during the post-questionnaire, so that participants in every condition answer on the same screen. A participant who closes the page partway can resume by opening the same link.

The recording component opens the avatar screen inside a frame and receives only in-page signals marking the start of capture, the end of an utterance, and the end of capture. The signals carry durations and nothing else. What reaches the server is the participant ID, the condition, usage time for each stage, the AI's speaking time and number of utterances, the questionnaire responses, the browser type, and a list of timestamped event types such as the start and end of capture and of each utterance; the audio, its features, and the content of the conversation do not. To remove stimuli unrelated to the experiment, participants are not shown experience points, the panels for saving and reviewing phrases, or the model switcher; the experiment instructions and questionnaires take their place. The API for saving and reading phrases is also disabled on the server side, and only the one configured model is served.

On the participant's side, what is needed is a desktop Chromium-based browser, speakers or earphones, and an account for the dialogue service. 

Many services offer voice dialogue only to logged-in users; with such a service, participants have to provide an account and log in. Whether a free account can use voice dialogue, and how far its usage time is limited, need to be checked before a study is run. The researcher's only cost is running the server; no usage fee proportional to the amount of dialogue arises. Providing an account may, however, be a barrier to recruitment in online experiments.

There are three caveats. First, which plans can use voice dialogue, the usage limits, and the model that serves it are decided by the provider and may change without notice. A dialogue stops partway if a limit is reached, so the length of the dialogue must be designed to fit within it. Second, participants' accounts differ in the plan subscribed to, the voice selected, the memory of past conversations, and so on. These become differences separate from the presentation condition. Randomisation keeps them from being unbalanced across conditions; in addition, the plan and similar attributes should be asked in the questionnaire and recorded together with the time of the study. Third, the conversation is sent to the provider under the participant's account and remains in that account's history. Using the server edition still adds no party that touches the conversation, but the fact that the conversation reaches the provider and remains in the history should be included in the explanation given to participants.

\section{Limitations and Future Work}
\label{sec:eval}

No user study has yet been conducted, and the effects of avatar presentation on human--AI interaction remain untested. The specific experimental design using the presentation conditions described in Section \ref{sec:instrument}, and its outcome variables, will be preregistered separately.

Deskmate has several limitations. First, where a rendering backend uses a third party's runtime or model asset, the applicable licensing and usage terms must be checked. This, however, is a constraint specific to the chosen backend, not a property of the side that generates the avatar state; in a configuration using the CSS backend, which requires no asset, or an open-source avatar, it does not arise. 
Second, the server edition of Section \ref{sec:server} is a prototype. It has no authentication or rate limiting, so anyone who knows a participant ID can submit (only the first submission per ID is accepted). Records are stored in a file on the server, and there is no mechanism for recovery from failure. These must be added before the edition is used in a full study. The desktop implementation does not support the server edition.

\section*{Beta Demo and Inquiries}

A beta demo of the browser implementation of Deskmate is available at the following page.
The demo includes a single VRM model. 

\begin{center}
\url{https://hiroshihamada.github.io/deskmate-web-beta/}
\end{center}

The demo lets one try tab-audio coupling and the 3D and CSS backends. The desktop implementation, the 2D backend, and the features for experiments are included in the full version, available from the author on request for research purposes only. Those interested in research or commercial use of Deskmate should contact the author.

The text and figures of this paper are licensed under CC BY 4.0; the software, including the demo, is not. Copyright in the software belongs solely to Hiroshi Hamada. The VRM model used in the demo and shown in the figures was made by the first author from VRoid Studio sample data. It falls under neither licence, and the terms set by pixiv Inc. extend to it.

\section*{Acknowledgements}

VRoid Studio was used to create the 3D model shown in the figures.

\section*{Disclosure of AI Use}

Generative AI assistance tools were used in part in preparing this paper and in implementing the software. The authors have verified and are responsible for the final text, the design decisions, the technical claims, and the decision to publish.

\end{document}